\documentclass[prl,showpacs,superscriptaddress,twocolumn,amsfonts,amsmath,floatfix]{revtex4}
\usepackage[sort&compress]{natbib}
\usepackage[final]{graphicx}
\newlength\figurewidth
\AtBeginDocument{\setlength\figurewidth{.9\linewidth}}
\let\vec\boldsymbol
\def\kT{\ensuremath{k_\text{B}T}}

\begin{document}
\title{Active and Nonlinear Microrheology in Dense Colloidal Suspensions}
\date{\today}
\def\mpdlr{\affiliation{%
  Institut f\"ur Materialphysik im Weltraum,
  Deutsches Zentrum f\"ur Luft- und Raumfahrt (DLR),
  51170 K\"oln, Germany}}
\def\unikn{\affiliation{%
  Fachbereich Physik, Universit\"at Konstanz,
  78457 Konstanz, Germany}}
\def\unial{\affiliation{%
  Departamento de F\'\i{}sica Aplicada, Universidad de Almer\'\i{}a,
  04.120 Almer\'\i{}a, Spain}}
\author{I.~Gazuz}\unikn
\author{A.~M.~Puertas}\unial
\author{Th.~Voigtmann}\unikn\mpdlr
\author{M.~Fuchs}\unikn

\begin{abstract}
We present a first-principles theory for the active nonlinear
microrheology of colloidal model systems: for constant
external force on a spherical probe particle embedded in a dense host
dispersion, neglecting hydrodynamic interactions, we derive an
exact expression for the friction. Within mode-coupling theory (MCT),
we discuss the threshold external force needed to delocalize the probe
from a host glass, and its relation to strong nonlinear
velocity-force curves in a host fluid.
Experimental microrheology data and simulations, which we
performed,
are explained with a simplified model.
\end{abstract}

\maketitle

Microrheology is a promising technique
providing local probes of the dynamics in a complex fluid \cite{Waigh.2005}.
Monitoring the motion of a singled-out probe particle embedded
in a (dense) host fluid or gel, one addresses
questions about the microscopic origins of the host's complex-fluid behavior
and in particular the link between microscopic
mechanisms and macroscopic properties amenable to conventional rheology.
This connection
subtly depends on the host, probe-bath
interactions, and on the applied forces.
Active microrheology turns this into an advantage, at the cost of
requiring much better knowledge about the microscopic processes
\cite{Squires.2008}: applying a known forcing to the
individual particle, one explores the nonequilibrium and usually nonlinear
regime, providing detailed insight into the structure-dynamics
relationship, e.g., in cellular environments \cite{Wilhelm.2008} or
close to the glass transition \cite{Hastings.2003,Habdas.2004,Williams.2006}.
Laser tweezers, magnetic or surface-chemistry forces \cite{Erbe}
provide experimental realizations achieving large forcing.

The external-force-velocity relations obtained in dense suspensions
reveal striking nonlinearities, induced by the slow relaxation of the
host. Leaving the linear-response regime,
a sudden strong increase in the velocity 
reveals the strength required to pull free the probe from the
(transient) local neighbor cage, such that force-induced motion overrules
structural relaxation.
Recent theoretical progress \cite{Squires.2005,Jack.2008} notwithstanding,
it remains to understand the nonlinear
friction induced by the slow structural rearrangements of host particles.

Here we develop a theory for active nonlinear microrheology
in suspensions close to their glass transition, establishing
the conceptual connection between micro- and macro-rheology, when
(de)localization of
the probe occurs, and how the structure of the cage is distorted close
to this yielding point.
We start from
microscopic equations of motion and relate the force--velocity relation
of the probe, by virtue of an exact Green-Kubo-like formula,
to a microscopic-force autocorrelation function. This
can be approximated through non-equilibrium tagged-particle density
correlation functions, which in turn are calculated
in the framework of
the mode-coupling theory of the glass transition (MCT) \cite{Goetze1992}.
For a hard-sphere (HS) suspension with a pulled probe
of same size as the host particles, we demonstrate
that the theory predicts a delocalization threshold force
that explains the nonlinear response seen in
experiment and simulation.

We start from the many-body
Smoluchowski equation for the nonequilibrium distribution function
$\Psi(t)$ of a system of $N$ Brownian particles
(positions $\vec r_i$) and a single
probe (labeled $s$),
$\partial_t\Psi(t)=\Omega\Psi(t)$.
Subjecting only the probe particle to a
constant, homogeneous force $\vec F^\text{ex}$, the Smoluchowski operator
$\Omega=\Omega_0+\Delta\Omega$ reads
\begin{equation}\label{eq:smol}
  \Omega = \sum_{i=1,\ldots,N,s}
    \vec\partial_{i}\cdot\left(k_BT\vec\partial_{i}-\vec
    F_{i}\right)/\zeta_i
  - 
  \left(\vec\partial_s\cdot\vec F^\text{ex}\right)/\zeta_s\,,
\end{equation}
where $\Delta\Omega=-(\vec\partial_s\cdot\vec F^\text{ex})/\zeta_s$
is the nonequilibrium term describing active forcing, and
$\Omega_0$ the equilibrium time-evolution. We
neglect solvent-induced hydrodynamic interactions and introduce
Stokes friction coefficients 
for host ($\zeta_{i=1,\ldots N}\equiv\zeta_0$) and probe ($\zeta_s$) particles.  The $\vec
F_{i,s}$ are (potential) interaction forces among the particles.

To obtain nonequilibrium averages formed with the force-dependent Smoluchowski
operator, the integration-through-transients
(ITT) formalism \cite{Fuchs2002c} recasts Eq.~\eqref{eq:smol}:
\begin{equation}\label{eq:dis}
  \Psi(t)=\Psi_\text{eq}
  -\frac\kT{\zeta_s} \int_0^t dt'\,\exp[\Omega t'](\vec F^\text{ex}\cdot
  \vec F_s)\Psi_\text{eq}\,,
\end{equation}
assuming equilibrium at $t=0$.
In particular, the stationary friction
coefficient $\zeta(F^\text{ex})$, defined via the average
stationary velocity at given external force
\begin{equation}\label{eq:zeta}
  \zeta\langle\vec v_s\rangle_{t\to\infty} \equiv
  \zeta\langle\vec v_s\rangle_\infty = \vec F^\text{ex}\,,
\end{equation}
is found by using Eq.~\eqref{eq:dis} and equating external and interaction
forces on the probe:
\begin{equation}\label{mueq}
  \zeta = \zeta_s+\frac{1}{3\kT}\int_0^\infty dt\,
  \langle\vec F_s\exp[\Omega^\dagger(\vec F^\text{ex})
  t]\vec F_s\rangle_\text{eq}\,.
\end{equation}
This formally exact generalized Green-Kubo relation connects the
far-from-equilibrium response to a transient
equilibrium-averaged correlation function.
ITT achieves that all following averages are equilibrium ones,
denoted by $\langle\cdot\rangle$
(suppressing the $\text{eq}$ subscript).
$\Omega^\dagger$ is the
adjoint of $\Omega$, and $\vec
F^\text{ex}$ enters non-perturbatively; linear response is recovered
by neglecting this dependence.

Following MCT, we assume that  force fluctuations are governed by
collective and probe-particle density fluctuations,
$\varrho_{\vec q}=\sum_{i=1}^N\exp[i\vec q\vec r_i]$ and
$\varrho^s_{\vec q}=\exp[i\vec q\vec r_s]$.
We take that in the
thermodynamic limit, the motion of the probe has negligible impact
on the bulk properties of the host, and restrict wave numbers
to a discrete grid neglecting anomalous long distance correlations.
Inserting a projector ${\mathcal P}_2\propto\sum_{\vec k\vec p}
\varrho^s_{\vec k}\varrho_{\vec p}\rangle
\langle\varrho^s_{\vec k}\varrho_{\vec p}$ on both sides of the
operator exponential in Eq.~\eqref{mueq}, because forces on the
probe relax by host particle rearrangements and probe motion, and
splitting four-point density averages into dynamical density
correlators, $\phi_{\vec k}(t)= \langle\varrho_{-\vec
k}\exp[\Omega^\dagger t]\varrho_{\vec k}\rangle$ and $\phi^s_{\vec
k}(t)= \langle\varrho^s_{-\vec k}\exp[\Omega^\dagger
t]\varrho^s_{\vec k}\rangle$, we arrive at
\begin{equation}\label{eq:ffmct}
  \langle\vec F_s\exp[\Omega^\dagger t]\vec F_s\rangle
  \approx \sum_{\vec k}\frac{|\kT kS_k^s|^2}{NS_k}\phi^s_{\vec k}(t)
  \phi_{-\vec k}(t)\,.
\end{equation}
$S_k=\langle\varrho_{\vec k}\varrho_{-\vec k}\rangle$ and
$S_k^s=\langle\varrho^s_{\vec k}\varrho_{-\vec k}\rangle$ are
the equilibrium structure functions describing interactions among
probe and host particles.

The probe correlator $\phi_{\vec q}^s(t)$
is complex-valued, as the perturbed operator $\Omega^\dagger$
is non-Hermitian. This reflects that the probe-density
distribution is shifted by application of an external force: while
in equilibrium it is centered around the origin, the average
position of the probe moves,
introducing a complex-valued phase factor in $\phi_{\vec q}^s(t)$.
Still, Eq.~\eqref{eq:ffmct} maintains $\zeta\in{\mathbb R}$ due to
the symmetry $\phi_{-\vec q}^s(t)=(\phi_{\vec q}^{s}(t))^*$.

Equation \eqref{eq:ffmct} recasts the problem of calculating the
probe friction as one of calculating collective and
tagged-particle density correlation functions.
To this end, we employ
Zwanzig-Mori equations of motion \cite{Fuchs2002c},
\begin{subequations}\label{eq:phis}
\begin{equation}\label{eq:mzphis}
  \partial_t\phi^s_{\vec q}(t)=-\omega^s_{\vec q,\vec q}\phi^s_{\vec q}(t)
  -\int_0^tdt'\,m^s_{\vec q}(t-t')\partial_{t'}\phi^s_{\vec q}(t')\,,
\end{equation}
closed by the MCT approximation generalizing Eq.~\eqref{eq:ffmct} to
finite wave vectors,
\begin{equation}\label{eq:mem}
  m^s_{\vec q}(t)=\frac\kT{\zeta_s \omega^s_{\vec q,\vec q}}
  \sum_{\vec k+\vec p=\vec q}
  \frac{1}{NS_p}
  {\mathcal V}_{\vec q\vec k\vec p}^s
  {\mathcal V}_{\vec q\vec k\vec p}^{s,\dagger}
  \phi^s_{\vec k}(t)\phi_{\vec p}(t)\,.
\end{equation}
\end{subequations}
Again, the
physical idea in the MCT approximation is that the friction kernel
$m^s_{\vec q}(t)$ relaxes by both probe and host density dynamics.
The coupling coefficients are ${\mathcal V}_{\vec q\vec k\vec p}^s=(\vec
q\vec p) S^s_p$, ${\mathcal V}_{\vec q\vec k\vec p}^{s,\dagger}=
  \omega^s_{\vec q,\vec p} S^s_p$, where   $\omega^s_{\vec q,\vec p}
  =(\vec q\kT-i\vec F^\text{ex})\cdot\vec p/\zeta_s$.
An analogous set of equations holds for $\phi_{\vec q}(t)$. Since
the external force acts on the probe only, the $\phi_{\vec q}(t)$
are in fact determined by the unperturbed Smoluchowski operator,
$\Omega_0$, resulting in the standard MCT scenario of glassy dynamics
\cite{Goetze1992,Megen1993}.
This describes arrest driven by wave vectors connected with a typical
host particle
radius $a$. Thus the dimensionless parameter measuring the effect of
the external force is $aF^\text{ex}/(\kT)$, the work required to pull the
probe over that distance in relation to thermal energy.

\begin{figure}
\includegraphics[width=.8\figurewidth]{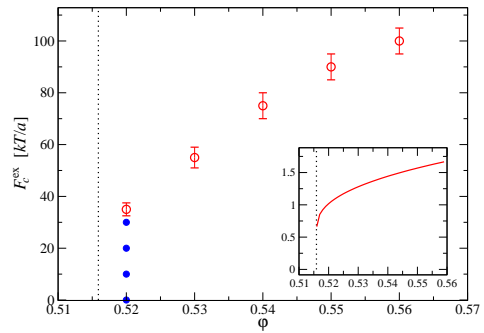}
\caption{\label{fig:fc-phi}
  Threshold force $F^\text{ex}_c(\varphi)$ needed to delocalize a
  hard-sphere probe particle in a glass of equally large hard spheres
  with packing fraction $\varphi$ above the glass transition (dotted line),
  calculated from MCT within the Percus-Yevick approximation.
  Blue circles mark $F^\text{ex}$ values used in Fig.~\ref{fig:contours}.
  Inset: corresponding schematic-model result (see text).
}
\end{figure}

The macroscopic counterpart to the friction $\zeta$ is the dispersion
viscosity $\eta$ measured in bulk flow. Within ITT, the analog to
Eq.~\eqref{mueq} holds
for the latter \cite{Brader2007}. MCT expresses this as a
functional only of the host correlators $\phi_{\vec q}(t)$, while in
Eq.~\eqref{eq:ffmct},
the probe correlators $\phi_{\vec q}^s(t)$ enter.
In linear response close to the glass transition,
identical scaling laws for both closely link
micro- and macro-rheology \cite{Fuchs1999c}.
For large external forces, this correspondence breaks:
Equations \eqref{eq:phis} for the probe correlator contain a novel
delocalization transition that is absent in $\phi_{\vec q}(t)$.
A probe arrested in a glassy host suspension remains localized in its
(deformed) nearest-neighbor cage (described by
$f^s_{\vec q}=\phi^s_{\vec q}(t\to\infty)>0$),
yielding zero average velocity
(infinite friction) only below a finite threshold
$F^\text{ex}_c$. At larger force,
the probe is pulled free ($f^s_{\vec q}=0$)
and attains a steady velocity (finite friction) at long times.
In the liquid, cages are transient, and a remnant of the threshold survives
as a sudden sharp ``force thinning'' in $\zeta(F^\text{ex})$.

The details of the delocalization transition depend on
the host properties, which we model now as hard spheres using the
known numerical
MCT results for the collective density correlators $\phi_{\vec q}(t)$
within the Percus-Yevick $S_q$-approximation
\cite{Franosch1997}. This model yields a glass transition at packing fraction
$\varphi_c\approx0.516$, where $\varphi=(4\pi/3)\varrho a^3$ with number
density $\varrho$ is the only parameter.
Figure~\ref{fig:fc-phi} shows our results for the
delocalization threshold force, $F^\text{ex}_c$, for a probe equal to
the host particles ($S_q=1+S^s_q$). This threshold
$F^\text{ex}_c(\varphi_c)>0$ is finite at $\varphi_c$, and increases
further with increasing density. Note that
$F^\text{ex}_c={\mathcal O}(50\,k_\text{B}T/a)$, much larger than one
might intuitively expect. This reflects the strong caging force
exerted by the set of nearest neighbors that must be overcome
before the probe can be delocalized.

\begin{figure}
\includegraphics[width=1.1\figurewidth]{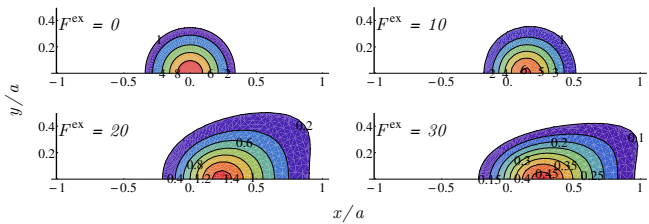}
\caption{\label{fig:contours}
  Contour plot of the
  probability distribution $f^s(\vec r)$ for a localized hard-sphere probe
  of radius $a$
  in a hard-sphere system with same radius
  at $\varphi=0.52$, for external forces acting
  to the right with indicated magnitude in units of $\kT/a$.
}
\end{figure}

The inverse Fourier transform of $f^s_{\vec q}$
is the $t\to\infty$ probability distribution
for the position of a probe starting at the origin;
Figure~\ref{fig:contours} shows our results
at packing fraction $\varphi=0.52$, slightly above
the glass transition, for several forces below
$F^\text{ex}_c$. $\vec F^\text{ex}$ is taken to be in
(positive) horizontal direction, rendering $f^s(\vec r)$
rotational-symmetric around this axis. For zero
force, the distribution is spherical-symmetric and centered around
the origin; 
it decays on a
length scale of $0.2 a$, the typical localization length
for solids dominated by hard-core repulsion. Small applied
forces mainly shift the center of the distribution to a position
$x_0\approx 0.2 a$, i.e., they push the probe to the ``cage wall''
without essentially distorting the cage. Close to the delocalization
threshold, however, $f^s(\vec r)$ develops a deformed tail extending into the
force direction,
reducing the spherical symmetry to a merely rotational
one. Interestingly, the tail does not extend along the
symmetry axis; rather, a ``dip'' is seen in direction of the
applied force.
For $F^\text{ex}\ge F^\text{ex}_c$, $f^s_{\vec q}$
and $f^s(\vec r)$ vanish, indicating a delocalized state.

To discuss probe friction or similar dynamical quantities,
we need to solve the time-dependent and spatially inhomogeneous
Eqs.~\eqref{eq:phis}. A first step towards this is to
solve a simplified, ``schematic''
MCT model. Take $\zeta=1+\int_0^\infty\,\phi^s(t)\phi(t)\,dt $, with
\begin{subequations}\label{schem}
\begin{gather}
  \partial_t\phi^s(t)+\omega^s\phi^s(t)+
  \int_0^t m^s(t-t')\partial_{t'}\phi^s(t')\,dt'=0\,,\\
  m^s(t)=v_s\phi^{s*}(t)\phi(t)\,,
\end{gather}
\end{subequations}
$\omega^s=1-iF^\text{ex}$. The host correlator $\phi(t)$
is set by
the ``F$_\text{12}$ model'' often used to describe glassy
dynamics in equilibrium \cite{Goetze1992,f12model},
governed by a separation parameter $\epsilon$ such that $\epsilon<0$
in the liquid, and $\epsilon\ge0$ in the glass; $\epsilon$ thus measures
the host interactions.
$v_s$ describes the strength of probe--host coupling.
Its $F^\text{ex}$ dependence is ignored,
neglecting the subtle interplay arising from
static correlations involving more than one wave vector;
still, the model recovers the qualitative behavior of the force threshold
(see inset of Fig.~\ref{fig:fc-phi}).
We expect the schematic model to be reasonable for small
enough velocities and/or forces,
where the friction is dominated by universal features
of the transition at $F^\text{ex}_c$.

\begin{figure}
\includegraphics[width=\figurewidth]{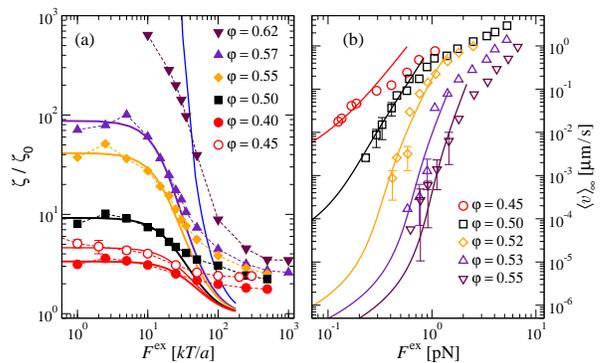}
\caption{\label{fig:sim}
  (a) Probe friction $\zeta$ as a function of external force for
  packing fractions $\varphi$ as indicated, from simulations of a
  quasi-hard-sphere system (symbols), from Brownian dynamics
  for monodisperse HS (Ref.~\cite{Carpen.2005}, open symbols),
  and from the schematic model
  (solid lines; see text) with $\kT/a=0.05586$, $1/\zeta_0=3.44$, and
  parameters
  $(\epsilon,v_s)=(0,6)$, $(-0.045,3.5)$, $(-0.058,3)$, $(-0.18,2)$,
  $(-0.5,1.5)$, $(-0.8,1)$, top to bottom.
  (b) Open symbols: Experimental force--velocity relations for a colloidal
  suspension, from Ref.~\cite{Habdas.2004}. Lines: schematic model with
  ($\epsilon,v_s)=(-0.006,46)$, $(-0.008,30)$, $(-0.01,15)$, $(-0.05,6)$,
  and $v_1=v_2=0.1=v_s/5$ (bottom to top), shifted along $x$ and $y$ by
  $0.058$ and $296$.
}
\end{figure}

To test the simplified model, we performed
simulations of a slightly polydisperse quasi-hard sphere system
undergoing strongly damped Newtonian dynamics, which shows a glass transition
at $\varphi_c\approx0.595$ \cite{Voigtmann2004b}. Particles
(mass $m=1$, $\kT=1$, radii distributed uniformly in $[0.9,1.1]$)
suffer friction with the solvent
($\zeta_0=50$) and random forces obeying the
fluctuation-dissipation theorem. One particle is randomly selected to undergo
an external force $\vec{F}^\text{ex}$ until it reaches a distance half the
size of the simulation box (elongated in the direction of
$\vec{F}^\text{ex}$ by a factor of $8$). The average probe velocity is
measured sampling more than 300 independent trajectories, and the friction is
calculated using Eq.~\eqref{eq:zeta}. All simulations were
initially equilibrated, except for $\varphi=0.62$, where the system was
aged for $t_w=25000$. At this density, results show little
influence of ageing for forces $F^\text{ex}\gtrsim35\,k_\text{B}T/a$.

A strong decrease in the dynamical friction $\zeta$ around
$F^\text{ex}_c={\mathcal O}(40\,\kT/a)$ seen in the simulation [symbols
in Fig.~\ref{fig:sim}(a)] indicates the force threshold.
Fitting $\epsilon$ and $v_s$ per curve, and two shift factors
setting the units, the schematic model reproduces this behavior for
$\varphi<\varphi_c$. In the (idealized) glass, it predicts a true
delocalization transition as $\zeta\to\infty$ for
$F^\text{ex}<F^\text{ex}_c$, exemplified by a $\epsilon=0$ curve in
Fig.~\ref{fig:sim}(a). In the simulation, $\zeta$ remains finite in the
accessible window presumably because of ergodicity restoring processes
ignored here \cite{Goetze1992}.
Employing larger $v_s$, our model also explains recent experiments on
colloidal systems using larger probes \cite{Habdas.2004}, as shown in
Fig.~\ref{fig:sim}(b). In these velocity-force curves obtained below
the glass transition, the force-threshold
signature is a steep increase of $\langle\vec v_s\rangle_\infty$
around $F^\text{ex}_c\approx0.2\,\text{pN}\gg \kT/a$, again reproduced by
the model.
For too large external forces, the model fails, as expected above.
%

\begin{figure}
\includegraphics[width=\figurewidth]{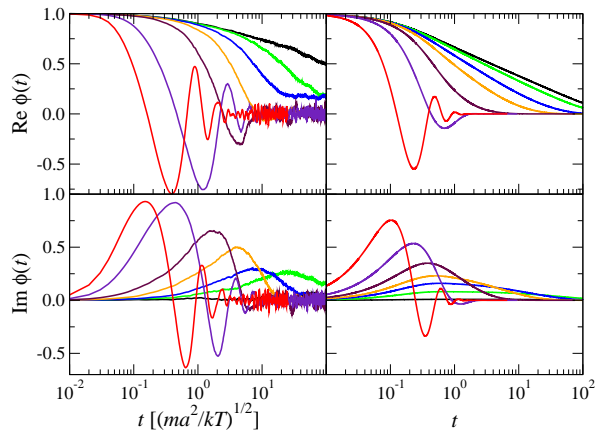}
\caption{\label{fig:fsqt}
  Probe-particle density correlation function
  $\phi_{\vec q}^s(t)$ from computer simulation at $\varphi=0.55$ (left) and
  from schematic MCT (right; fit as in Fig.~\ref{fig:sim}),
  real (top) and imaginary (bottom) parts. $\vec q\parallel\vec F^\text{ex}$
  corresponds to the
  position of the main peak in the static structure factor $S(q)$.
  For the simulation, $aF^\text{ex}/(\kT)=1$, $10$, $20$, $30$, $50$,
  $100$, and $250$ (right to left).
}
\end{figure}

The virtue of the schematic model is to allow more detailed qualitative
predictions for the slow nonequilibrium dynamics. This is
demonstrated by Fig.~\ref{fig:fsqt}, where we compare the tagged-particle
density correlation function $\phi^s_{\vec q}(t)$ obtained from the
simulation for a wave vector $\vec q\parallel\vec F^\text{ex}$ with a
magnitude
corresponding to the nearest-neighbor peak in $S_q$.
The simulation confirms the existence
of complex-valued correlation functions (for this $\vec q$-direction)
as a signature of non-equilibrium, naturally arising in our
microscopic framework. For $\vec q$ perpendicular to the external force,
$\phi^s_{\vec q}(t)=(\phi^s_{-\vec q}(t))^*$ remains real-valued, owing to the
rotational symmetry $\phi^s_{\vec q}(t)=\phi^s_{-\vec q}(t)$, and
exhibits the two-step decay typical for glass formers with an intermediate
plateau and a final relaxation sped up by the external force.
No such clear plateau is seen in the figure for $\vec q$ parallel to
$\vec F^\text{ex}$. For large forces, the $\phi^s_{\vec q}(t)$
show pronounced oscillations,
quite unexpected for a Brownian system, and even stronger in the
simulation data.

To summarize, we have developed a microscopically founded theory for
the nonlinear active microrheology close to a glass transition. Starting
from the Smoluchowski equation without hydrodynamic interactions, and
applying approximations in the spirit of the mode-coupling theory of the
glass transition, we predict the probe friction as a function of the
external force and of the equilibrium host structure.

The theory predics a finite microrheological force threshold needed to
delocalize a probe from a glassy host, locally melting it.
In the dense liquid, this is reflected by a strong nonlinear decrease
in friction coefficients differentiating the regimes where cages are
either broken by slow structural relaxation (for small external force),
or by large enough applied force.
A schematic model captures these aspects and allows to fit
experimental and simulation data quantitatively for not too large external
forces.

The force threshold could be related to the existence of
a yield stress well established for glassy colloidal systems, and predicted
by MCT for constant-velocity bulk driving.
It will be promising to study more closely
this relation and the dynamical behavior of
the system close to micro- and macro-yielding.

\begin{acknowledgments}
We thank A.~Erbe, W.C.K.~Poon and J.F.~Brady for discussions, and
Deutsche Forschungsgemeinschaft (SFB~513 project B12),
Helmholtz-Gemeinschaft (Hochschul-Nachwuchsgruppe VH-NG~406),
M.E.C.\ project MAT-2006-13646-CO3-02 and
Junta de Andaluc\'\i{}a (P06-FQM-01869)
for funding.
\end{acknowledgments}

\bibliographystyle{apsrev}
\bibliography{lit}

\begin{thebibliography}{18}
\bibitem[{f12()}]{f12model}
\bibinfo{note}{We use Eq.~\protect\eqref{schem} with $\omega=1$ for the
  correlator $\phi(t)$ and memory kernel $m(t)=v_1\phi(t)+v_2\phi(t)^2$ with
  $v_{1,2}=v_{1,2}^c(1+\epsilon)$ and $(v_1^c,v_2^c)=(0.828\ldots,2)$.}
\end{thebibliography}

\end{document}